\documentclass[aps,prl,preprint,nopacs,superscriptaddress]{revtex4}

\usepackage{graphicx}
\usepackage{verbatim}
\usepackage{mathrsfs}
\usepackage{amsmath,amsfonts,amssymb}
\usepackage{graphicx}
\def\3{2.8in}    %used for figure widths
\def\2{2.5in}
\def\4{3.0in}

\def \beq {\begin{equation}}
\def \eeq {\end{equation}}
\begin{document}

\title{Saddle point singularity and topological phase diagram in a tunable topological crystalline insulator}

%Coexistence of topological order and saddle point singularity in the phase diagram of a topological crystalline insulator

%Saddle point singularity and topological phase diagram of a topological crystalline insulator

%Saddle point singularity and topological phase diagram of a topological crystalline insulator
%Lifshitz transition and van Hove singularity on the surface of symmetry protected topological insulator} 
%Electronic phase diagram of topological crystalline insulator state in Pb$_{1-x}$Sn$_x$Se}
%Temperature, doping and crystal symmetry driven topological crystalline insulator phase in Pb$_{1-x}$Sn$_x$Se}
\author{Madhab Neupane*}\affiliation {Joseph Henry Laboratory, Department of Physics, Princeton University, Princeton, New Jersey 08544, USA}

\affiliation{Princeton Center for Complex Materials, Princeton University, Princeton, New Jersey 08544, USA}

\author{Su-Yang Xu*}\affiliation {Joseph Henry Laboratory, Department of Physics, Princeton University, Princeton, New Jersey 08544, USA}

\author{R. Sankar} \affiliation{Center for Condensed Matter Sciences, National Taiwan University, Taipei 10617, Taiwan}

\author{Q. Gibson}\affiliation {Department of Chemistry, Princeton University, Princeton, New Jersey 08544, USA}

\author{Y. J. Wang}\affiliation {Department of Physics, Northeastern University, Boston, Massachusetts 02115, USA}
\affiliation {Advanced Light Source, Lawrence Berkeley National Laboratory, Berkeley, California 94305, USA}

\author{N. Alidoust}\affiliation {Joseph Henry Laboratory, Department of Physics, Princeton University, Princeton, New Jersey 08544, USA}

%\author{Q. Gibson}\affiliation {Department of Chemistry, Princeton University, Princeton, New Jersey 08544, USA}

\author{G. Bian}\affiliation {Joseph Henry Laboratory, Department of Physics, Princeton University, Princeton, New Jersey 08544, USA}

\author{Chang Liu}\affiliation {Joseph Henry Laboratory, Department of Physics, Princeton University, Princeton, New Jersey 08544, USA}

\author{I. Belopolski}\affiliation {Joseph Henry Laboratory, Department of Physics, Princeton University, Princeton, New Jersey 08544, USA}

%\author{A.~Fedorov}\affiliation {Advanced Light Source, Lawrence Berkeley National Laboratory, Berkeley, California 94305, USA}
%\author{S.-K.~Mo}\affiliation {Advanced Light Source, Lawrence Berkeley National Laboratory, Berkeley, California 94305, USA}

%\author{J. D. Denlinger}\affiliation {Advanced Light Source, Lawrence Berkeley National Laboratory, Berkeley, California 94305, USA}
%\author{Y. J. Wang}\affiliation {Department of Physics, Northeastern University, Boston, Massachusetts 02115, USA}

\author{Y. Ohtsubo}\affiliation {Synchrotron SOLEIL, Saint-Aubin-BP 48, F-91192 Gif sur Yvette, France}

\author{A. Taleb-Ibrahimi} \affiliation {Synchrotron SOLEIL, Saint-Aubin-BP 48, F-91192 Gif sur Yvette, France} \affiliation {UR1/CNRSSynchrotron SOLEIL, Saint-Aubin, F-91192 Gif sur Yvette, France}

\author{S. Basak}\affiliation {Department of Physics, Northeastern University, Boston, Massachusetts 02115, USA}

\author{W.-F. Tsai}\affiliation {Department of Physics, National Sun Yat-sen University, Kaohsiung 80424, Taiwan}

\author{H. Lin}\affiliation {Graphene Research Centre and Department of Physics, National University of Singapore, Singapore 117542}
%\author{L. A. Wray}\affiliation {Joseph Henry Laboratory, Department of Physics, Princeton University, Princeton, New Jersey 08544, USA}\affiliation {Advanced Light Source, Lawrence Berkeley National Laboratory, Berkeley, California 94305, USA}

%\author{R. Sankar} \affiliation{Center for Condensed Matter Sciences, National Taiwan University, Taipei 10617, Taiwan}

\author{R. J. Cava}\affiliation {Department of Chemistry, Princeton University, Princeton, New Jersey 08544, USA}

\author{A. Bansil}\affiliation {Department of Physics, Northeastern University, Boston, Massachusetts 02115, USA}

%\author{R. Sankar} \affiliation{Center for Condensed Matter Sciences, National Taiwan University, Taipei 10617, Taiwan}

\author{F. C. Chou} \affiliation{Center for Condensed Matter Sciences, National Taiwan University, Taipei 10617, Taiwan}

%\author{A. Bansil}\affiliation {Department of Physics, Northeastern University, Boston, Massachusetts 02115, USA}

\author{M. Z. Hasan}
\affiliation {Joseph Henry Laboratory, Department of Physics, Princeton University, Princeton, New Jersey 08544, USA}
\affiliation{Princeton Center for Complex Materials, Princeton University, Princeton, New Jersey 08544, USA}


\begin{abstract}
\textbf{A topological crystalline insulator (TCI) is a new phase of topological matter, which is predicted to exhibit distinct topological quantum phenomena, since space group symmetries replace the role of time-reversal symmetry in the much-studied Z$_2$ topological insulators \cite{graphene, graphene_th, graphene_Kondo,graphene_super, kim, Geim, RMP, Zhang_RMP}. Utilizing high-resolution angle-resolved photoemission spectroscopy (ARPES), we reveal the momentum space nature of interconnectivity of the Fermi surface pockets leading to a saddle point singularity within the topological surface state alone in the TCI Pb$_{0.7}$Sn$_{0.3}$Se.
%we report the direct observation of topological surface band structure saddle point on the surface of a topological crystalline insulator Pb$_{0.7}$Sn$_{0.3}$Se. 
%The surface saddle point is observed at the location in the energy and momentum space corresponding to the Lifshitz transition, where near each $\bar{\textrm{X}}$ point, two unconnected surface Fermi surfaces are found to merge into two concentric pockets. 
Moreover, we show that the measured momentum-integrated density of states exhibits pronounced peaks at the saddle point energies, demonstrating the van Hove singularities (VHSs) in the topological surface states, whose surface chemical potential, as we show, can be tuned via surface chemical gating, providing access to the topological correlated physics on the surface. Our experimental data reveal a delicate relationship among lattice constant, band gap and spin-orbit coupling strength associated with the topological phase transition in Pb$_{1-x}$Sn$_{x}$Se. Furthermore, we explore the robustness of the TCI phase with VHS in Pb$_{1-x}$Sn$_{x}$Se, which shows a variety of distinct topological phase transitions driven by either thermal instability or broken crystalline symmetry, and thus revealing a rich topological phase diagram  in Pb$_{1-x}$Sn$_{x}$Se for the first time.}

%by systematically exploring the robustness of the topological crystalline insulator phase in Pb$_{1-x}$Sn$_{x}$Se system, we observe various topological phase transitions due to thermal instability or structural transition, where a previously undiscovered rich electronic phase diagram in Pb$_{1-x}$Sn$_{x}$Se is found. Our observation of topological order and saddle point singularity pave the way for realizing various novel states emerging through coupling to symmetry breaking interactions on the topological surfaces of the tunable Pb$_{1-x}$Sn$_{x}$Se system.}

\end{abstract}
\date{\today}
\maketitle

Saddle point singularities in two-dimensional Dirac electron systems often result in electronic instabilities leading to exotic correlated quantum phenomena \cite{graphene, graphene_th}. A notable example is the
observation of novel phenomena associated with the 
 van Hove singularity (VHS) in graphene, where exotic superconductivity, unusual magnetism or the Kondo effect have been predicted to take place \cite{graphene, graphene_th, graphene_Kondo,graphene_super}, all of which is intimately connected to the VHS in the electronic structure. Very recently, striking HofstadterÕs butterfly spectrum and fractional quantum Hall effect have been experimentally observed in twisted graphene systems due to the saddle point singularity in its Dirac band structure \cite{kim, Geim}. Like graphene, the surface states of  Z$_2$ topological insulators (TIs) also possess light-like Dirac dispersion \cite{RMP, Zhang_RMP} but they are spin-momentum locked as a consequence of the nontrivial topology of the bulk band structure \cite{RMP, Zhang_RMP}. As a result, saddle point singularities in topological surface states (a spin-helical two-dimensional Dirac electron gas) are even more desirable because it can not only host correlated physics, but also allows one to explore the coexistence and interplay between topological order and strong correlation. However, to date, saddle point singularity in Z$_2$ TIs has not been realized. This is because a band structure saddle point usually comes from a change of Fermi surface topology, which requires multiple Fermi surfaces to be interconnected in some specific way in energy-momentum space. Unfortunately, surface states of the Z$_2$ TIs are odd in number which must locate encircling the KramerÕs points in the surface Brillouin zone. For example, well-known Z$_2$ TIs such as Bi$_2$Se$_3$, Bi$_2$Te$_3$ and TlBiSe$_2$ feature single-Dirac cone surface state that have only one surface state Fermi pocket enclosing the $\Gamma$ point  \cite{RMP, Zhang_RMP}. Therefore, realizations of saddle point singularities in the surface states of Z$_2$ TIs are essentially difficult.
 
 A relaxation of the  Z$_2$ condition can also lead to topological surface state which is the case for the new topological crystalline insulator (TCI) phase. In  the TCI phase, the topological order arises from the crystalline space group symmetry  \cite{Liang NC SnTe, Liang PRL TCI}. The  surface states are not thus confined to enclose the KramerÕs points, therefore it is possible to have a multiple topological surface Fermi surfaces intersect each other which is a critical condition for realizing a saddle point singularity \cite{Ando, Suyang, PSS TCI, Lifshitz}. In this paper, we present the first observation of \textit{momentum-resolved} electronic structure topology providing a direct view of the saddle point singularity in the surface state band dispersion of the TCI Pb$_{0.70}$Sn$_{0.30}$Se.
 Previously, scanning tunneling microscopy (STM) has been used to study graphene and TCI systems \cite{graphene, stm}, where an enhanced density of states ($dI/dV$) peak at a certain energy has been interpreted to be associated with the observation of a van Hove singularity (VHS). However, existence of a weak peak in the density of states (DOS) does not shed light regarding its origin. Moreover, a VHS is defined in momentum space through intersecting FS pockets connected in a very specific way which can not be resolved without a momentum space probe. The momentum structure of the bands that leads to singularity is critical in understanding and providing the type of correlation electron behavior that can potentially be realized in this system.
% VHS is defined in momentum space intersecting Fermi surfaces which can not be resolved without a momentum space probe. Moreover, the momentum structure of the bands that lead to singularity is critical in understanding and providing the type of correlation electron behavior that can be related in this system.
In this paper, we present critical ARPES measurements along with supporting model calculations for the Pb$_{1-x}$Sn$_x$Se system.
Our results reveal the momentum space nature of the interconnectivity of the FS pockets leading to a saddle point  singularity within the topological surface state alone. Combining high-quality crystal with surface gating, we prepare the topological surface state which made the unambiguous identification of the singularity possible for the first time. We further show that  the existence of the singularity can be tuned through a thermodynamic phase transition.
%to experimentally show 1. momentum-resolved surface Fermi surface Lifshitz transition, 2. momentum-space surface band saddle points at the Lifshitz transition energy, 3. enhanced momentum-integrated ARPES DOS at the corresponding Lifshitz transition and saddle point energy, 4. exclusion of bulk band origin.
 Our experiments with the critically important momentum resolution, collectively, provide a view of topological state hosting a VHS, its origin from band structure saddle point, and its thermodynamic tunability involving the topological surface states only in Pb$_{0.70}$Sn$_{0.30}$Se, which is not possible in Bi$_2$Se$_3$ class of TI. Furthermore, we explore the robustness of the TCI phase with VHS in Pb$_{1-x}$Sn$_x$Se, which shows a variety of distinct topological phase transitions driven by either thermal instability or broken crystalline symmetry, and thus revealing a rich topological phase diagram in Pb$_{1-x}$Sn$_x$Se material which was previously unknown. Our observation of a saddle point singularity on the topological surface of Pb$_{0.70}$Sn$_{0.30}$Se opens the door for testing the interplay between topological order and strong electron-electron correlation phenomena, which can potentially host exotic superconductivity \cite{Kopaev}, topological Kondo insulator \cite{Coleman}, and other correlated topological phases \cite{Balents, Moore_2}.

In order to identify for possible saddle point singularity in Pb$_{0.7}$Sn$_{0.3}$Se, we systematically study its surface state electronic structure, presented in Figs. 1 and 2. The energy evolution of the surface state constant energy contour is shown in Fig. 2c. Two Dirac points are observed to be located at the same energy $E_{\textrm{B}}=70$ meV (central panel of Fig. 2c) near the $\bar{X}$ point along the $\bar{\Gamma}-\bar{X}-\bar{\Gamma}$ momentum space direction in the surface Brillouin zone (BZ). As we move away from the Dirac point in energy, the two Dirac points grow into two unconnected pockets, and eventually ``meet'' each other, where they are found to become two concentric contours both enclosing the $\bar{X}$ point. Therefore a surface state Lifshitz transition is observed in our data since the constant energy contour is found to undergo a topological change. We now focus on the constant energy contour at the Lifshitz transition energy at $E_{\textrm{B}}=40$ meV (Fig. 1a). The two dots (green and blue) in Fig. 1a mark the momentum space locations, where the two unconnected contours merge. Their energy and momentum space coordinates are experimentally identified to be $( E_{\textrm{B}}, k_x, k_y)=(40$ meV$, 0, \pm0.02$ $\textrm{\AA}^{-1})$. To experimentally prove the saddle points in momentum-space band structure, we focus on the upper blue dot in Fig. 1a and study the energy-momentum dispersion cuts along three important momentum space cut-directions, namely cuts 1, 2, and 3. Cuts 1 and 2 (Fig. 1c and d) are cut along the horizontal ($k_x$) and vertical ($k_y$) directions across the blue dot. Interestingly, the blue dot is found to be a local band structure minimum along cut 1 shown in Fig. 1c, whereas it is a local maximum along cut 2 (Fig. 1d). Observation of a local minimum and local maximum at the same momentum space location (the blue dot) manifestly shows that it is a surface band structure saddle point. Such observation is uniquely enabled by our momentum-resolved (ARPES) probe, which is not possible with other experimental techniques such as STM or transport. The observation of surface momentum-space saddle point immediately implies that there exists certain intermediate cut-directions (between cuts 1 and 2), where the surface band structure is completely flat in the vicinity of the blue dot. Indeed, as shown in Fig. 1e, for cut 3, we found that the surface states are nearly flat near the saddle point.

The observed flat band structure (along cut 3) may give rise to the divergence of the surface density of states (DOS), leading to a surface VHS. 
%We first present the theoretically calculated momentum-integrated DOS of the TCI surface states. As shown in Fig. 2c, two DOS peaks are found, which correspond to the VHSs in the upper and lower parts of the Dirac cones, respectively. 
There are two peaks in the theoretically calculated momentum-integrated DOS for the TCI surface states, Fig. 2c. These correspond to the VHSs in the upper and lower parts of the Dirac cones.
Additionally, three dips in the calculated DOS curve correspond to the two Dirac points, as well as the upper and lower Dirac points (UDP and LDP) (see schematics in Fig. 2b). In order to search for experimental evidence of VHSs, we study the momentum-integrated ARPES intensity as shown in Fig. 2f. Indeed,  a pronounced peak is observed at the energy corresponding to the saddle point, namely $E_{\textrm{B}}=40$ meV, as labeled by ``VH1'' in Fig. 2f. Additionally, we observe a significant dip of ARPES intensity at the binding energy of $E_{\textrm{B}}=70$ meV, which corresponds to the energy of the Dirac points. Other theoretically expected features, including the VHS of the lower part of the Dirac cones (``VH2'') as well as the UDP and the LDP are not significant in the raw momentum-integrated ARPES data. This is very likely due to the fact that at the corresponding energies the bulk valence (conduction) bands appear to overlap with the surface states in energy and momentum space, which can significantly weaken the cross-section of the surface states (e.g. the case in Fig. 2d at $k\sim0$ and $E_{\textrm{B}}>0.1$ eV) due to surface resonance effects. In order to better highlight those features, we take second derivative of the ARPES intensity (Fig. 2f right), in which now all five features are visualized.  On the other hand, the ARPES intensity peak observed at $E_{\textrm{B}}=40$ meV (``VH1'') is very clear, since at that energy the surface states are isolated from the bulk bands and hence ARPES dispersion is sharp and distinct as seen in Fig. 2d. In order to further exclude the bulk origins of the observed peak at $E_{\textrm{B}}=40$ meV, we show calculated bulk DOS in the right panel of Fig. 2c. The bulk DOS is zero inside the band-gap and monotonically increases as the energy is moved away from the bulk band-gap. No intensity peaks are found in a $\pm0.05$ eV window. Thus we conclude the observed ARPES intensity peak at $E_{\textrm{B}}=40$ meV originates from the topological surface states. The observation of photoemission intensity peak at $E_{\textrm{B}}=40$ meV provides strong evidence to the theoretically predicted van Hove singularity. Such data are analogous to the enhanced $dI/dV$ near the Lifshitz transition energy observed in STM measurements in graphene or Pb$_{0.70}$Sn$_{0.30}$Se. %which has been attributed to the direct observation of van Hove singularity. 
However, we emphasize that observation of enhanced density of states \textit{alone without momentum resolution} is not sufficient to prove the case of a singularity and furthermore its physical origin, since the experimentally observed DOS does not diverge and other extrinsic reasons such as enhanced DOS from appearance of bulk bands or even impurity bands cannot be ruled out without momentum resolution. Here, our experimental observation of 1. a momentum-resolved surface Fermi surface Lifshitz transition, 2. a momentum-space surface band saddle points at the Lifshitz transition energy, 3. an enhanced momentum-integrated ARPES DOS at the corresponding Lifshitz transition and saddle point energy, 4. the exclusion of bulk band origin, collectively, provide a comprehensive and convincing experimental proof of a van Hove singularity and also its origin from band structure saddle point in the topological surface states in Pb$_{0.70}$Sn$_{0.30}$Se. 
Furthermore, chemical potential tuning by Sn deposition (Fig. 2a) on the surface of TCI provides the unambiguous route for direct identification of the singularity in this system. 

%\bigskip
%\bigskip
%\textbf{Results}
%\newline
%\textbf{Temperature evolution of single gapped state from two Dirac cone like dispersion}
%\newline

%Fig. 4a shows the Fermi surface plot for Pb$_{0.70}$Sn$_{0.30}$Se measured at temperature (T) of 20 K with photon energy of 18 eV. 

In order to study the fundamental properties of the TCI phase and the construction of potential topological devices, it is crucial to demonstrate the tunability between the TCI phase and the topologically trivial phase via multiple tuning parameters. Therefore, we study various topological phase transitions in the Pb$_{1-x}$Sn$_x$Se system and the systematic evolution of the surface electronic structure through the transitions. ARPES dispersion maps of Pb$_{0.70}$Sn$_{0.30}$Se at different temperatures are shown in Fig. 3a. At low temperatures of $T=20$ K, two Dirac cones are observed near the $\bar{X}$ point along the $\bar{\Gamma}-\bar{X}-\bar{\Gamma}$ cut, which is evident in the nontrivial TCI phase. Interestingly, as temperature is raised to 100K, the two Dirac points are found to move closer to each other. When further increasing the temperature, the two cones are observed to merge at temperature around 250 K at the $\bar{X}$ point. And finally a gap at the Dirac point is opened for $T>250$ K, which suggests that the system enters the topologically trivial phase. To understand the interesting thermodynamic evolution of the surface state electronic structure, we perform first principles theoretical calculations at different lattice constant values (Fig. 3b). A reasonably good agreement between the temperature dependent ARPES data and the lattice constant dependent calculation is found, where the essential features of the data, including the two Dirac cones approaching, merging and eventually opening up a gap, are all captured in the calculation. Thus our ARPES data and calculation results together suggest that the observed thermally driven topological phase transition is a result of thermal expansion of the lattice. To further confirm that, we perform synchrotron-based X-ray diffraction measurements on the sample as a function of temperature. As shown in Fig. 3c, the peak of the diffraction $2\theta$ angle is clearly found to move towards smaller values with rising temperature, which is qualitatively consistent with the picture of thermodynamic expansion of the lattice constant (see supplementary information (SI) for details). 
%However, quantitatively, it can be seen that from the XRD data the lattice is found to expand for maximumly $1\%$, whereas calculation predicts an increase of more than $3\%$ in order to drive the system from TCI to trivial phase. This seems to suggest that the observed thermal evolution is more complicated than a simple lattice expansion origin and some other factors can also play an important role in the thermal topological transition. 

We now compare our observation with the topological phase transition found in the Z$_2$ TI system in BiTl(S$_{1-\delta}$Se$_{\delta}$)$_2$ \cite{Hasan QPT} to highlight interesting properties of the topological phase transition in Pb$_{0.70}$Sn$_{0.30}$Se. First, we observe that the momentum-space distance between the two Dirac points near each $\bar{X}$ point in Pb$_{0.70}$Sn$_{0.30}$Se can be systematically engineered. We note that changing the surface Dirac point momentum space location is allowed in a mirror protected TCI system \cite{Liang NC SnTe}, as long as the Dirac points are on the mirror lines. This is, in contrast, not possible in BiTl(S$_{1-\delta}$Se$_{\delta}$)$_2$ since it has only single Dirac cone at the surface BZ center (a time-reversal invariant Kramer's point) as required by the time-reversal protected TI phase. Manipulation of the momentum space distance between the two Dirac points is of importance, because a longer distance between the two Dirac points means a larger saddle point curvature. This consequently leads to a more pronounced VHS (more states populate at the VH energy), which therefore can be used as a knob to tune the surface state correlation strength. Second, at the topological phase transition critical point in Pb$_{0.70}$Sn$_{0.30}$Se, the two Dirac cones merge and also the bulk band-gap vanishes. Thus both the surface and the bulk are characterized by a Dirac dispersion, and the system can be viewed as a \textit{square-version of graphene} in 3D, potentially leading to exotic transport and tunneling behaviors waiting to be explored \cite{Ong}. 
Third, it is also interesting to note that a similar temperature dependence study on BiTl(S$_{0.4}$Se$_{0.6}$) ($\delta=0.6$ corresponds to a composition in the TI phase but close to the critical point (see Ref.\cite{Hasan QPT}) does not show any dramatic temperature dependent effect, which is in sharp contrast to the clear temperature driven band inversion and topological transition observed in Pb$_{0.70}$Sn$_{0.30}$Se. This is very likely due to the fact that Pb$_{0.70}$Sn$_{0.30}$Se is in the cubic crystal structure so there exists only one parameter $a$ for the lattice constant. On the other hand, the layered van der Waals structure in BiTl(S$_{1-\delta}$Se$_{\delta}$) allows at least two free parameters (in plane $a$ and out-of-plane $c$), which can have opposite effects to the band inversion as a function temperature that cancel each other. Furthermore, our systematic temperature dependent surface electronic structure evolution through the topological transition in Pb$_{0.70}$Sn$_{0.30}$Se makes contrasts with previous study on a similar system Pb$_{0.77}$Sn$_{0.23}$Se \cite{PSS TCI}, where the two Dirac cones do not merge but directly open a gap as temperature is raised. Here, our data and supporting calculations seem to suggest a much clearer ``lattice thermal expansion origin'' for the observed topological transition.

We study the TCI phase in Pb$_{1-x}$Sn$_x$Se as a function of composition $x$. Fig. 4a shows the ARPES measurements of the low energy states of the $x=0.3$ sample, as well as the two end compounds, namely PbSe and SnSe. For PbSe, low-lying bulk conduction and valence bands with a clear band-gap of $\sim0.15$ eV is observed, which suggests that the system is topologically trivial for $x=0$. As $x$ are increased (see the SI for $x=0.18$ and $x=0.23$ data), the low-lying bulk bands are observed to approach each other, and eventually inverse with the surface states spanning over the inverted band-gap. The band inversion critical composition is found to be $\sim0.23$ depending on the temperature. We note that the fact that the increasing Sn concentration drives the system from trivial to TCI has also been found in the Pb$_{1-x}$Sn$_x$Te. This is interesting since Pb is in fact heavier than Sn thus having a larger (atomic) spin-orbit strength. However, in both Pb$_{1-x}$Sn$_x$Se and Pb$_{1-x}$Sn$_x$Te, shrinkage of lattice constant (with increasing Sn\%) plays a more important role and therefore makes the system topologically nontrivial in the Sn-rich side. We turn to the other end compound SnSe, as shown in the third panel in Fig. 4a. Surprisingly, at the Fermi level, no electronic states are observed. Instead, a fully gapped electronic structure with the chemical potential inside the band-gap is found. This is because the SnSe crystal is in the orthorhmobic phase \cite{SnSe Crystal, SnSe_distor}. The change of crystal structure, which breaks both inversion and mirror symmetry, destroys the TCI phase.
%,which denies both the band inversion and the mirror symmetries, completely destroys the TCI phase. 
And SnSe is found to be a trivial insulator by our ARPES measurements and the band-gap is reported to be as large as $\sim1$ eV \cite{SnSe_gap}. Based on our systematic ARPES studies, a rich topological phase diagram is found in the Pb$_{1-x}$Sn$_x$Se in Fig. 4b. The blue and red lines represent the energy level of the lowest lying bulk conduction and valence bands assuming that the crystal structure always remains in the FCC structure. Starting from PbSe ($x=0$), the system has a non-inverted band-gap of $\sim0.15$ eV. As $x$ increases, band inversion takes place and the system enters the TCI phase. The inverted band-gap increases until the system enters the multi-(crystal structure)-phase regime at $x\gtrsim0.45$, where cubic and orthorhombic structures coexist. Finally, for $x\gtrsim0.75$, the system becomes a large band-gap trivial insulator in the single orthorhombic phase. Two distinct phase transitions are observed (as labeled by $x_{\textrm{c1}}$ and  $x_{\textrm{c2}}$), where the first transition is due the the shrink of lattice constant (which increases the \textit{effective} spin-orbit strength), whereas the second is a result of a drastic structural transition. Therefore, our experimental data reveal a delicate relationship among lattice constant, band gap, spin-orbit coupling strength and crystal structure associated with the topological phase transition in Pb$_{1-x}$Sn$_{x}$Se.

\bigskip
\bigskip
\textbf{Methods}
\newline

\textbf{Experiment}

Single crystals of Pb$_{1-x}$Sn$_{x}$Se used in these measurements were grown by standard growth method (see ref.\cite{Growth_1, SnSe_Stru_2} and supplementary informations for detailed sample growth). Low photon energy (15 eV-30 eV) and temperature dependent ARPES measurements for the low-lying electronic structures were performed at the Synchrotron Radiation Center (SRC) in Wisconsin and the Stanford Synchrotron Radiation Centre (SSRL) in California with R4000 electron analyzers whereas high photon energy ($\sim$60 eV) measurements were performed at the Beamlines 12 and 10 at the Advanced Light Source (ALS) in California, equipped with high efficiency VG-Scienta SES2002 and R4000 electron analyzers. Sn deposition ARPES measurements were done in The CASSIOPEE beamline, Soleil, France.
Samples were cleaved {\it in situ} and measured at 10-300 K in vacuum better than $1\times10^{-10}$ torr. They were found to be very stable and without degradation for the typical measurement period of 20 hours. 

Powder X-ray diffraction was performed on a Bruker D8 Focus x-ray diffractometer operating with CuK$\alpha$ radiation and a graphite diffracted beam monochromator. Resistivity measurements were performed on a Quantum Design Physical Property Measurement System (PPMS). 

\bigskip
\textbf{Theoretical calculations}

\bigskip
\textbf{First-principles calculations}
The first-principles calculations are carried out within the framework of the density functional theory (DFT) 
%using pseudo-potential projected augmented wave method 
using projector augmented wave method \cite{paw} as implemented in the VASP package \cite{vasp}. The generalized gradient approximation (GGA) \cite{gga} is used to model exchange-correlation effects. The spin orbital coupling (SOC) is included in the self-consistent cycles. The cutoff energy 260 eV is used for both PbSe and SnSe systems. To capture the physical essence of the surface state electronic structure of the TCI phase in an alloy Pb$_{1-x}$Sn$_{x}$Se system, we take SnSe and assume that SnSe crystalizes in the FCC structure and then calculate the $(001)$ surface electronic structure. We also calculate the electronic structure of PbSe (FCC) and SnSe (orthorhombic) using experimentally determined crystal structure and lattice constant (shown in the SI).  For all calculations, the $(001)$ surface is modeled by periodically repeated slabs of 48-atomic-layer thickness with  
24 $\textrm{\AA}$ vacuum regions and use a 12 $\times$ 12 $\times$ 1 Monkhorst-Pack  $k-$point mesh over the (BZ). 

\bigskip
\textbf{$k\cdot p$ model}
We use an effective surface $k\cdot p$ Hamiltonian \cite{kdotp} of topological crystalline insulator on $(001)$ plane to obtain the surface bands dispersion and spin texture of Pb$_{0.7}$Sn$_{0.3}$Se (see detailed in the SI and also Ref. \cite{kdotp}).

\newpage

\noindent

\newpage

\begin{figure*}
\centering
\includegraphics[width=17cm]{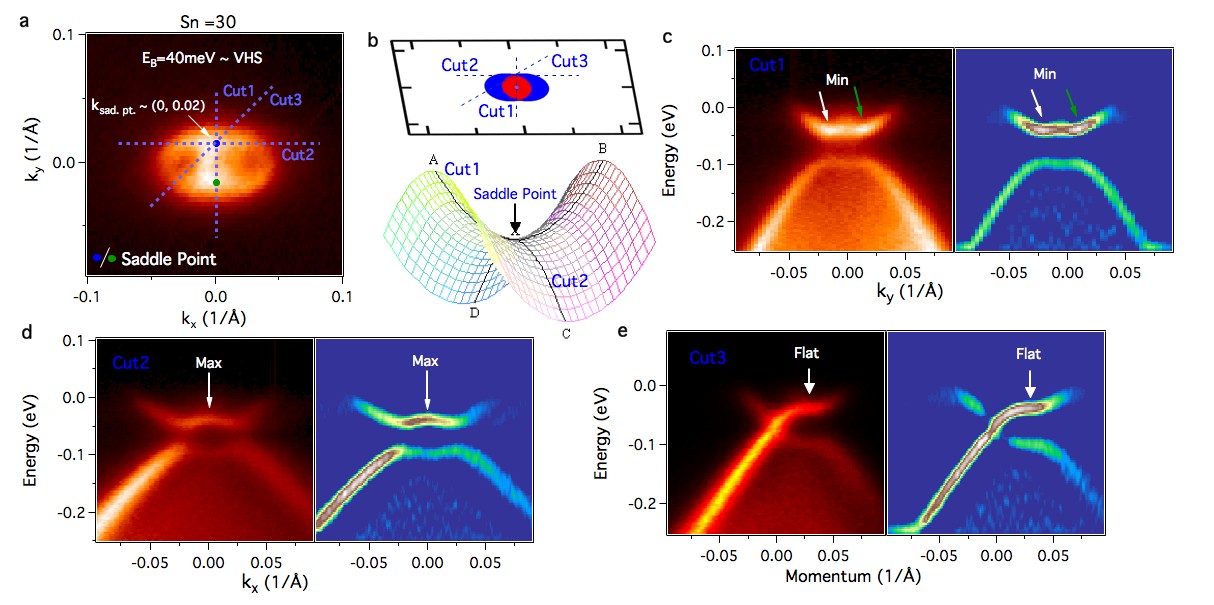}
\caption{\label{Fig1}\textbf{Observation of the saddle point curvature.} \textbf{a,} ARPES constant energy contour map in the vicinity of an $\bar{\textrm{X}}$ point in the surface (001) BZ at binding energy 40 meV, which corresponds to the surface Lifshitz transition and saddle point energy of the upper part of the Dirac cones. The $\bar{\textrm{X}}$ points locate at $(k_x,k_y)=(\pm0.7$ $\textrm{\AA}^{-1},0)$ or $(0, \pm0.7$ $\textrm{\AA}^{-1})$. Since the low-energy physics including topological surface states occur at the $\bar{\textrm{X}}$ points, we re-define the momentum space coordinate of one $\bar{\textrm{X}}$ point to be $(k_x,k_y)=(0,0)$ for simplicity of presentation. The blue and green dots denote the momentum space locations of the two surface saddle points. The blue dotted lines indicates the momentum space cut-directions for cuts1, 2, and 3, which are centered at the blue dot. (b) Calculated surface state constant energy contour at the saddle point singularity energy (top) and a three-dimensional schematic of a saddle point (bottom). \textbf{c-e,} ARPES dispersion maps (left) and their second derivative images along cuts 1, 2, and 3. The white and green arrow point the saddle points (blue and green dots in Panel \textbf{a}).}
\end{figure*}

\begin{figure*}
\centering
\includegraphics[width=17cm]{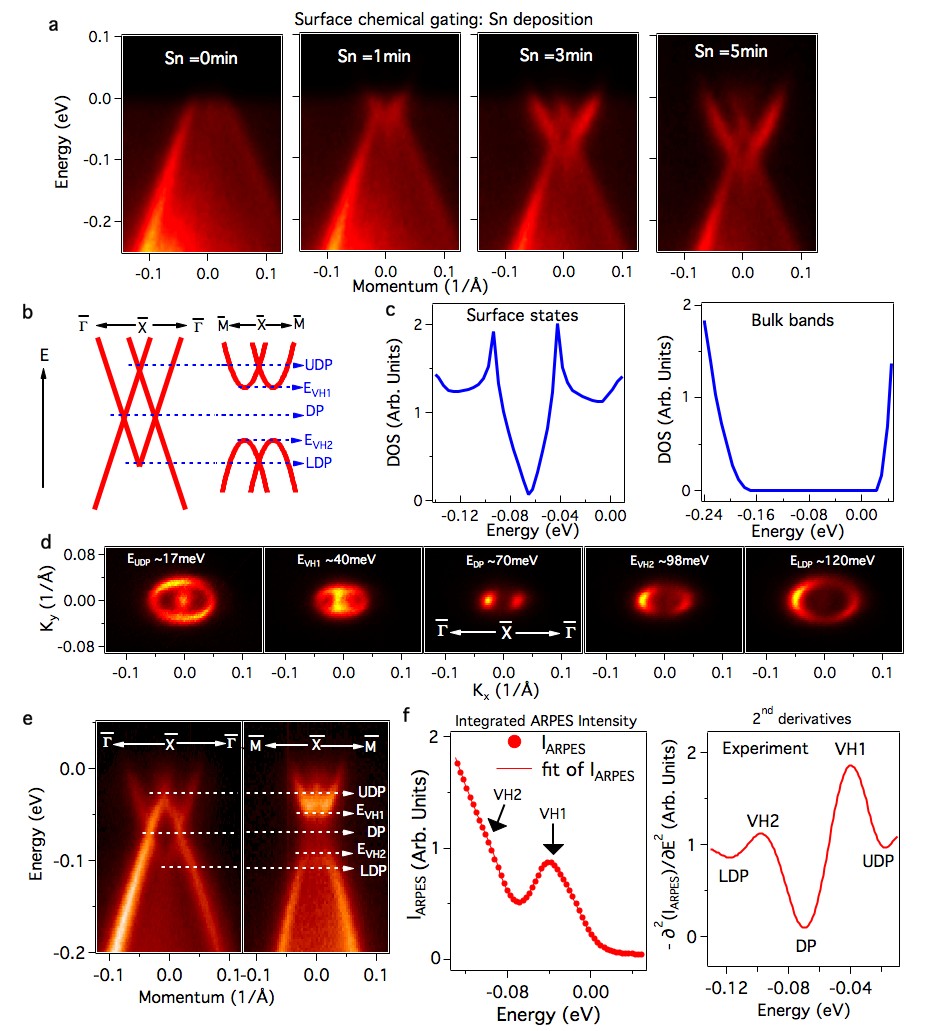}
\caption{\label{Fig2}\textbf{Observation of the saddle point singularity.} \textbf{a,} ARPES dispersion maps upon \textit{in situ} Sn deposition on the Pb$_{0.70}$Sn$_{0.30}$Se surface. The dosage (time) for Sn deposition is noted. A different batch of sample, which is $p-$type with the chemical potential below the Dirac points, is used for the Sn deposition data shown in this panel. \textbf{b and c,} Schematics of surface band dispersion of the TCI phase along the mirror line $\bar\Gamma-\bar{\textrm{X}}-\bar\Gamma$ and the $\bar{\textrm{M}}-\bar{\textrm{X}}-\bar{\textrm{M}}$ momentum space cut-directions.}
\end{figure*}

\addtocounter{figure}{-1}
\begin{figure} [t!]
\caption{Five important features of the surface states, including Dirac point of the upper part of the Dirac cones (UDP), van Hove singularity of the upper Dirac cones (VH1), two Dirac points along the $\bar\Gamma-\bar{\textrm{X}}-\bar\Gamma$ mirror line (DP), van Hove singularity of the lower part of the Dirac cones (VH2) and Dirac point of the lower part of the Dirac cones (LDP) are marked. \textbf{c,} Calculated density of state (DOS) for the surface states and the bulk bands using the $k\cdot p$ model \cite{kdotp}. \textbf{d,} Experimental observation of the Lifshitz transition - the binding energies are noted on the constant energy contours. \textbf{e,} ARPES measured dispersion plots along $\bar\Gamma-\bar{\textrm{X}}-\bar\Gamma$ and $\bar{\textrm{M}}-\bar{\textrm{X}}-\bar{\textrm{M}}$.  \textbf{f,} Momentum ($k_x$ and $k_y$) integrated ARPES intensity as a function of binding energy (left). 2$^{nd}$ derivative of the ARPES intensity with respect to binding energy is presented to further highlight the features. The upper Dirac point (UDP), upper van Hove singularity (VH1), Dirac point (DP), lower van Hove singularity (VH2) and lower Dirac point (LDP) are marked.}
\end{figure}

\clearpage
\begin{figure}
\centering
\includegraphics[width=17cm]{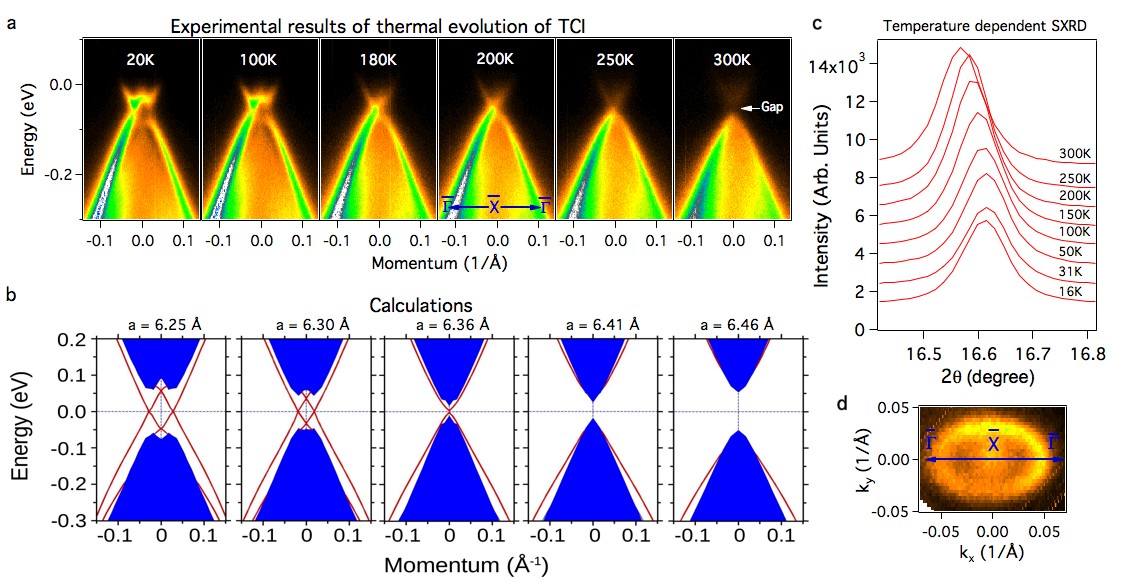}
\caption{\label{Fig3} \textbf{Thermal evolution of surface states topology (TCI $\rightarrow$ Trivial)}
\textbf{a,} Dispersion maps along the mirror line $\bar\Gamma-\bar{\textrm{X}}-\bar\Gamma$ at different temperatures. The two Dirac points are observed to approach, eventually merge into one node and then open up a gap as temperature increases. \textbf{b,} First-principles calculations of the TCI surface bands with varying lattice constants. SnSe assumed in the face-center cubic (FCC) structure is used. Red lines and blue areas represent the surface and bulk bands, respectively. \textbf{c,} Synchrotron-based temperature dependent X-ray diffraction (SXRD) measurements for Pb$_{0.70}$Sn$_{0.30}$Se. The peak is observed to shifts towards the lower angles with increasing temperature, which confirms the picture of the thermal expansion of lattice (see supplementary information for details). \textbf{d,} Surface Fermi surface plot measured at temperature of 20 K.} 
\end{figure}

\begin{figure*}
\centering
\includegraphics[width=16cm]{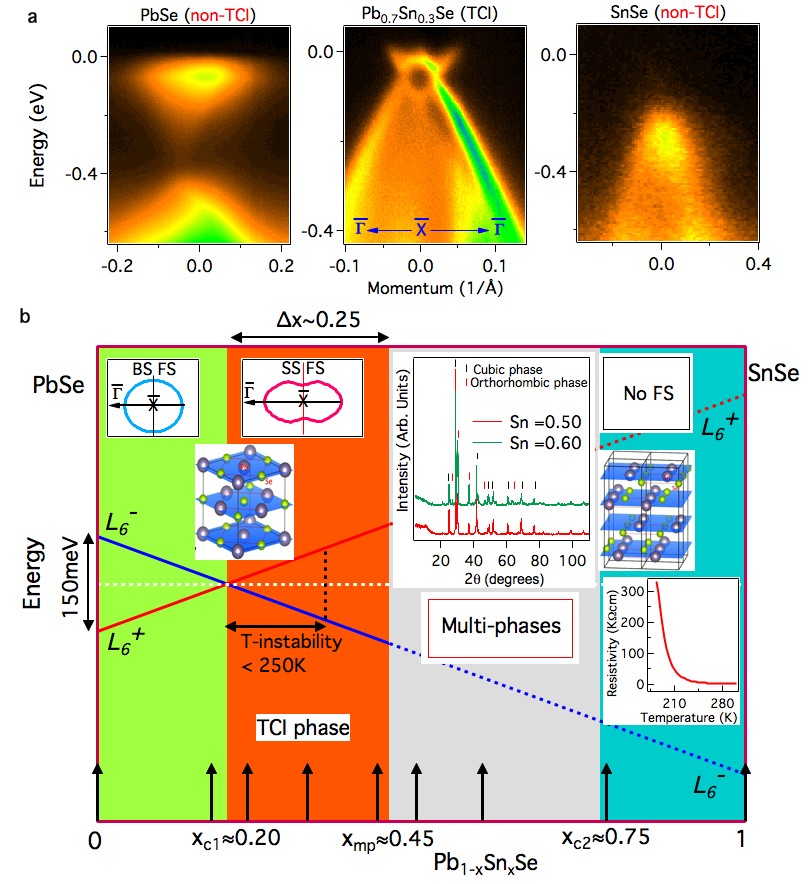}
\caption{\label{Fig4}\textbf{Topological phase diagram Pb$_{1-x}$Sn$_{x}$Se.} \textbf{a,} ARPES dispersion map for PbSe (trivial insulator), Pb$_{0.7}$Sn$_{0.3}$Se (TCI) and SnSe (trivial insulator). \textbf{b,} Topological phase diagram of the Pb$_{1-x}$Sn$_{x}$Se system. For composition range of $0<x<0.45$, the system is in the single crystalline FCC phase. The bulk band of Pb$_{1-x}$Sn$_{x}$Se undergoes a band inversion with Pb/Sn substitution. Topological crystalline insulator (TCI) phase is observed in the band inverted region toward the Sn-rich side. The critical composition $x_{c1}$ is $\sim0.20-0.23$ depending on the temperature. The conduction and valence band states representing odd and even parity eigenvalues are marked as }
\end{figure*}

\addtocounter{figure}{-1}
\begin{figure} [t!]
\caption{$L_6^-$ and $L_6^+$, respectively. For composition range of $0.45<x<0.75$, the system shows multi-structural-phase (cubic and orthorhombic phases coexist. See the XRD data in the inset for $x=0.5$ and $0.6$.). The upper insets are schematic Fermi surface plots around the $\bar{\textrm{X}}$ point. The inset in the bottom right conner shows the resistivity measurements on SnSe, which proves its insulator nature. BS FS and SS FS denote the bulk state Fermi surface and surface state Fermi surface, respectively. The arrows at bottom note the compositions where our ARPES studies have been performed. For composition range of $0.75<x<1$, the system is in a single crystalline orthorhombic phase.}
\end{figure}


\begin{thebibliography}{21} 

\bibitem{graphene} Li, G. \textit{et al.} Observation of van Hove singularities in twisted graphene layers. \textit{Nature Phys.} $\mathbf{6}$, 109-113 (2010).

\bibitem{graphene_th}Nilsson, J. \textit{et al.} Electron-electron interactions and the phase diagram of a graphene bilayer. \textit{Phys. Rev. B} $\mathbf{73}$, 214418 (2006) .


\bibitem{graphene_super} Nandkishore, R., Levitov, R., L. S \& Chubukov, A. V. Chiral superconductivity from repulsive interactions in doped graphene.  \textit{Nature Phys.} $\mathbf{8}$, 158-163 (2012).


\bibitem{graphene_Kondo}Lipinski, S., Krychowski, D.  Kondo effect near the van Hove singularity in biased bilayer graphene. 	arXiv:1206.4455 (2013)



\bibitem{kim} Dean, R. C. \textit{et al}. Hofstadter's butterfly in moire superlattices: A fractal quantum Hall effect. \textit{Nature} $\mathbf{497}$, 598-602 (2013).


\bibitem{Geim} L. A. Ponomarenko, A. L. \textit{et al}. Cloning of Dirac fermions in graphene superlattices. \textit{Nature} doi:10.1038/nature12187, (2013).




%\bibitem{Moore} Moore, J. E. The birth of topological insulators. \textit{Nature} $\mathbf{464}$, 194-198 (2010).

\bibitem{RMP} Hasan, M. Z. \& Kane, C. L. Colloquium: topological insulators.  \textit{Rev. Mod. Phys.} $\mathbf{82}$, 3045-3067 (2010).

\bibitem{Zhang_RMP} Qi, X-L. \& Zhang, S-C. Topological insulators and superconductors.  \textit{Rev. Mod. Phys.} $\mathbf{83}$, 1057-1110 (2011).


%\bibitem{David_nature} Hsieh, D. \textit{et al}. A topological Dirac insulator in a quantum spin Hall phase.  \textit{Nature} $\mathbf{452}$, 970-974 (2008).


%\bibitem{Xia} Xia, Y. \textit{et al}. Observation of a large-gap topological-insulator class with a single Dirac cone on the surface.  \textit{Nature Phys.} $\mathbf{5}$, 398-402 (2009).


%\bibitem{Chen_Science} Chen, Y. L. \textit{et al}. Experimental Realization of a Three-Dimensional Topological Insulator, Bi$_2$Te$_3$. \textit{Science} $\mathbf{325}$, 178-181 (2009).

%\bibitem{David_Science} Hsieh, D. \textit{et al}. Observation of Unconventional Quantum Spin Textures in Topological Insulators. \textit{Science} $\mathbf{323}$, 919-922 (2009).


%\bibitem{YLChen} Chen, Y.-L. \textit{et al}. Massive Dirac Fermion on the surface of a magnetically doped topological insulator.  \textit{Science} $\mathbf{329}$, 659-662 (2010).


\bibitem{Liang PRL TCI} Fu, L. Topological Crystalline Insulators. \textit{Phys. Rev. Lett.} $\mathbf{106}$, 106802 (2011).

\bibitem{Liang NC SnTe} Hsieh, H. \textit{et al}. Topological Crystalline Insulators in the SnTe Material Class. \textit{Nature Communications} $\mathbf{3}$, 982 (2012).

%\bibitem{PST Rhombohedral Distortion} Iizumi, M. \textit{et al}. Phase Transition in SnTe with Low Carrier Concentration. \textit{J. Phys. Soc. Jpn.} $\mathbf{38}$, 443-449 (1975).


\bibitem{Ando}Tanaka, Y. \textit{et al.} Experimental realization of a topological crystalline insulator in SnTe. \textit{Nature Phys.} $\mathbf{8}$, 800-803 (2012).

\bibitem{Suyang} Xu. S.-Y. \textit{et al.} Observation of a topological crystalline insulator phase and topological phase transition in Pb$_{1-x}$Sn$_x$Te. \textit{Nature Communications} $\mathbf{3}$, 1192 (2012).

\bibitem{PSS TCI} Dziawa, P. \textit{et al.} Topological crystalline insulator states
in Pb$_{1-x}$Sn$_x$Se. \textit{Nat. Material} $\mathbf{11}$, 1023-1027 (2012).

%\bibitem{PSS_TCI_spin} Wojek, B. M. \textit{et al.} Spin-polarized (001) surface states of the topological crystalline insulator Pb$_{0.73}$Sn$_0.27$Se. \textit{Phy. Rev B} $\mathbf{87}$, 115106 (2013).

%\bibitem{Ando_PST_2}Tanaka, Y. \textit{et al.} Tunability of the k-space Location of the Dirac Cones in the Topological Crystalline Insulator Pb$_{1-x}$Sn$_x$Te. arXiv:1301.1177 (2013).



\bibitem{Lifshitz} Lifshitz, I. M. \textit{et al.} Anomalies of electron characteristics of a metal in the high pressure region. \textit{Sov. Phys. JETP} $\mathbf{11}$, 1130 (1960).

\bibitem{stm} Okada, Y. \textit{et al.} Observation of Dirac node formation and mass acquisition in a topological crystalline insulator. \textit{Science} $\mathbf{27}$, 1496-1499 (2013).


\bibitem{Kopaev}  Belyavsky, V. I. Kapaev, V. V.  \& Kopaev, Y. V. Topological d-wave superconductor. \textit{JETP Lett} $\mathbf{96}$, 724-729 (2013).



\bibitem{Coleman} Dzero, M. \textit{et al.} Topological Kondo Insulators. \textit{Phys. Rev. Lett.} $\mathbf{104}$, 106408 (2010).




\bibitem{Moore_2} Mong, R. S. K. Essin, A. M. \& Moore, J. E. Antiferromagnetic topological insulators. \textit{Phys. Rev. B} $\mathbf{81}$, 245209 (2010). 





\bibitem{Balents} Pesin, D. \& Balents, L. Mott physics and band topology in materials with strong spinÐorbit interaction. \textit{Nature Phys} $\mathbf{6}$, 376-381 (2010).



%\bibitem{SnTe p-type} Burke, Jr., J. R., Allgaier, R. S., Houston, Jr., B. B., Babiskin, J. \& Siebenmann, P. G. Shubnikov-de Haas effect in SnTe. \textit{Phys. Rev. Lett.} $\mathbf{14}$, 360-361 (1965).



%\bibitem{SnSe_stru} Strauss, A. J. Inversion of Conduction and Valence Bands in Pb$_{1-x}$Sn$_x$Se Alloys. \textit{Phys. Rev. B} $\mathbf{157}$, 608-611 (1967).


\bibitem{Hasan QPT} Xu, S. Y. \textit{et al.} Topological Phase Transition and Texture Inversion in a Tunable Topological Insulator. \textit{Science} $\mathbf{332}$, 560-564 (2011).


\bibitem{Ong} Liang, T. \textit{et al.} Evidence for massive bulk Dirac Fermions in Pb$_{1-x}$Sn$_x$Se from Nernst and thermopower experiments. arxiv.org/abs/1307.4022 (2013).

%\bibitem{Chen Science BiTe} Chen, Y. L. \textit{et al}. Experimental Realization of a Three-Dimensional Topological Insulator, Bi$_2$Te$_3$. \textit{Science} $\mathbf{325}$, 178-181 (2009).


%\bibitem{SnSe_growth}Yu, J. G. \textit{et al.} Growth and electronic properties of the SnSe semiconductor. \textit{J. Crystal. Growth} $\mathbf{54}$, 248-252 (1981). 

%\bibitem{Growth} Nugraha, K. \textit{et. al.}  Yokota, growth and electrical properties of PbTe bulk crystals grown by the Bridgman method under controlled tellurium or lead vapor pressure. \textit{J. Cryst. Gr.} $\mathbf{165}$, 402-407 (1996).


%\bibitem{SnTe p-type} Burke, Jr., J. R., Allgaier, R. S., Houston, Jr., B. B., Babiskin, J. \& Siebenmann, P. G. Shubnikov-de Haas effect in SnTe. \textit{Phys. Rev. Lett.} $\mathbf{14}$, 360-361 (1965).

%\bibitem{SnTe ARPES} Littlewood, P. B. \textit{et al}. Band structure of SnTe studied by Photoemission Spectroscopy. \textit{Phys. Rev. Lett.} $\mathbf{105}$, 086404 (2010).


%\bibitem{Suyang_preformed}Xu, S.-Y. \textit{et. al.}, Anomalous spin-momentum locked two-dimensional states in the vicinity of a topological phase transition. arXiv:1204.6518 (2012).

%\bibitem{PbSe_gap}Lent, C. S. \textit{et. al.} Relativistic empirical tight-binding theory of the energy bands of GeTe, SnTe, PbTe, PbSe, PbS, and their alloys. \textit{Superlattices Microstruct.} $\mathbf{2}$, 491-499(1986).



\bibitem{SnSe Crystal} Okazaki, A. \& Ueda, I. The Crystal Structure of Stannous Selenide SnSe. \textit{J. Phys. Soc. Japan} $\mathbf{11}$, 470-470 (1956).

\bibitem{SnSe_distor}Volykhov, A. A. \textit{et. al.} Phase relations between germanium, tin, and lead chalcogenides in pseudobinary systems containing orthorhombic phases. \textit{Inorganic Materials} $\mathbf{44}$, 345-356 (2008).

\bibitem{SnSe_gap} Albers, W. \textit{et. al.} Preparation and properties of mixed crystals SnS$_{(1-x)}$Se$_x$. \textit{J. Phys. Chem. Solids} $\mathbf{23}$, 215-220 (1962).


\bibitem{SnSe_Stru_2} Szczerbakow, A. \& Berger, H. Investigation of the composition of vapour-grown Pb$_{1-x}$Sn$_x$Se crystals ($x\leq0.4$) by means of lattice parameter measurements. \textit{J. Crystal Growth} $\mathbf{139}$, 172-178 (1994).





\bibitem{Growth_1} Szczerbakow, A. \& Durose, K. Self-selecting vapour growth of bulk crystals Ð Principles 
and applicability. \textit{Prog. Cryst. Growth Charact. Mater.} $\mathbf{51}$, 81Ð108 (2005). 




%\bibitem{Growth_2} ] Szczerbakow, A. \& Berger, H. Investigation of the composition of vapour-grown Pb$_{1-x}$Sn$_x$Se crystals ($x \leq 0.4$) by means of lattice parameter measurements. \textit{J. Crystal Growth} $\mathbf{139}$, 172Ð178 (1994). 

\bibitem{paw} 
Kresse, G. \& Joubert, D. From ultrasoft pseudopotentials to the projector augmented-wave method. \textit{Phys. Rev. B} {\bf 59}, 1758 (1999).
 
\bibitem{vasp} 
Perdew, J. P., Burke, K., Ernzerhof, M. Generalized gradient approximation made simple. \textit{Phys. Rev. Lett.} \textbf{77}, 3865-3868 (1996).

 
\bibitem{gga} 
Kresse, G. \& Hafner, J.  Ab initio molecular dynamics for open-shell transition metals. \textit{Phys. Rev. B} {\bf 48}, 13115 (1993).


\bibitem{kdotp} Wang, Y. J. \textit{et al.} Nontrivial spin texture of the coaxial Dirac cones on the surface of topological crystalline insulator SnTe. \textit{Phys. Rev. B} $\mathbf{87}$, 235317 (2013).


 
 
%\bibitem{SnSe-Orth-cry}  Krebs, H., Langner, D., Mischkristallsysteme zwischen halbleitenden Chalkogeniden der vierten Hauptgruppe. II,  \textit{Z. Anorg. Allg. Chem.} $\mathbf{334}$, 37 (1964). 
 
%\bibitem{Ref_KP} Wang, Y. J. \textit{et al.} Nontrivial spin texture of the coaxial Dirac cones on the surface of topological crystalline insulator SnTe. \textit{Phy. Rev. B} $\mathbf{87}$, 235317 (2013).


%\bibitem{PSS TCI Preprint} Dziawa, P. \textit{et al}. \textit{arXiv:1206.1705} (2012).

\end{thebibliography}
\end{document}